# Title: Mixed equilibrium/nonequilibrium effects govern surface mobility in polymer glasses

**Authors:** Jianquan Xu[1], Asieh Ghanekarade[2], Li Li[1], Huifeng Zhu[1], Hailin Yuan[3], Jinsong Yan[3], David S. Simmons[2]*, Ophelia K. C. Tsui[3,4]*, Xinping Wang[1]*

**Affiliations:**

[1]School of Chemistry and Chemical Engineering, Key Laboratory of Surface & Interface Science of Polymer Materials of Zhejiang Province, Zhejiang Sci-Tech University, Hangzhou 310018, China.

[2]Department of Chemical, Biological, and Materials Engineering, University of South Florida, Tampa, Florida 33620, United States.

[3]Department of Physics, Hong Kong University of Science and Technology, Hong Kong Special Administrative Region 999077, China.

[4]William Mong Institute of Nano Science and Technology, Hong Kong University of Science and Technology, Hong Kong Special Administrative Region 999077, China.

*Corresponding authors. Emails: dssimmons@usf.edu (DSS); okctsui@ust.hk (OKCT); and wxinping@zstu.edu.cn (XW).

**Abstract:** The temperature at which supercooled liquids turn into solid-like glasses ($T_g$) can change at the free surface, affecting the properties of nanostructured glasses and their applications. However, inadequate experimental resolution to determine the $T_g$ gradient and a longstanding debate over the role of nonequilibrium effects have hindered fundamental understanding of this phenomenon. Using spatially resolved $T_g$ measurements and molecular dynamics simulations, we reveal a crossover from equilibrium behavior to a new regime of near-surface nonequilibrium glass physics on cooling. This crossover causes the form of the nonequilibrium $T_g$ gradient to change, highlighting the need to include these physics for rational understanding of the properties of realistic nanostructured glass-forming materials. They also potentially recast the interpretation of decades of experimental data on nanoconfined glasses.

## Introduction

Since the 1990s (*1, 2*), it has been well-established that the glass transition temperatures ($T_g$) of organic and polymer glass-formers change upon confinement in nano-scale domain size ($h$). These alterations have garnered significant attention. On the one hand, they play a central role in determining diverse properties of nanostructured materials, such as viscosity (*3*), dielectric response (*4*), thermal expansivity (*5, 6*), and heat capacity (*7*). On the other hand, the observation of broken-symmetry or finite-size alterations in the glass transition has long been anticipated to shed light on the nature of glass formation itself—a longstanding problem in condensed matter physics. However, it remains unsettled how the surface-induced alterations in dynamics and $T_g$ propagate into the material over distances of ~10 nm (*8, 9*) to ~1 μm (*10-12*), which is far greater than the propagation of alterations in thermodynamics and structural properties (*8, 9*).

Over the past decade, molecular dynamics (MD) simulations have revealed a coherent picture of alterations in dynamics near interfaces under equilibrium conditions. This understanding suggests that the relaxation time ($\tau$) of glasses exhibit a double exponential gradient over ~10 nm from the surface (*9, 13-16*), and the associated equilibrium dynamical $T_g$, denoted as $T_g^{eq}$, takes on an exponential gradient form (*15*). Experimental confirmation of these gradient has been suggested as a critical indicator of the nanoconfinement effect and could potentially reveal insights into the mechanism of the glass transition (*15*). However, there have been concerns about whether these gradient forms obtained from simulations (which have relatively short timescales of ≤ ~1 μs) persist at experimental timescales (which are typically ~100 s at the glass transition). But several results suggest that this should be a significant issue. For instance, simulations by Kob et al. (*13*) and Hocky et al. (*14*) showed that the double-exponential mobility gradient saturates rather than continuing to evolve with increasing simulation times. Recent experiments by Li et al. (*17*) and Zhang et al. (*18*) also support the double-exponential form of the gradient of surface diffusivity in organic glasses and relaxation time in a colloidal glass, respectively. To confirm if the exponential form of $T_g$ gradient found in simulations is also present in experiment, we conducted measurements using three surface techniques with different probe depths to determine the local $T_g$ as a function of depth ($z$) in various polymers. Our results show that the $T_g(z)$ gradients are linear in $z$. It is important to note that the experiments by Li et al. (*17*) and Zhang et al. (*18*) were conducted at temperatures ($T$) above the bulk $T_g$ ($T_{g,bulk}$), while ours were conducted at or below $T_{g,bulk}$, making our experiments more susceptible to nonequilibrium effects.

Experimental studies typically define $T_g$ by observing a gain (or loss) of equilibrium during heating (or cooling). Since these studies necessarily involve the system in a nonequilibrium state before (or after) the glass transition, we refer to this $T_g$ as the nonequilibrium $T_g$ ($T_g^{ne}$). In contrast, typical simulations as discussed above (*8, 9, 13, 15*) and some experiments (*3, 4, 19*) measure viscosity or $\tau$ at a fixed temperature under equilibrium or during a slow temperature scan (*20*). These studies investigate equilibrium dynamics and the equilibrium $T_g^{eq}$ and are commonly used to gain insights into the physics of glass transition. The difference between $T_g^{ne}$ and $T_g^{eq}$ in terms of the involvement of a nonequilibrium state during the glass transition have sparked a long-standing debate on whether nonequilibrium effects may impact common experimental measurements of $T_g$ (*20, 21*). This highlights the potential necessity of considering nonequilibrium effects when interpreting experimental results.

To gain a better understanding of our experimental result and the potential influence of nonequilibrium effects on the glassy dynamics of nanoconfined glasses in general, we perform MD simulations on the near-surface dynamics of long-chain polymers, incorporating *extrinsic* nonequilibrium effects, associated with an initially nonequilibrium chain conformational state, and *intrinsic* nonequilibrium effects, which refer to the unavoidable loss of equilibrium at $T_g^{ne}$, to the systems. The results from these simulations not only explain the observed $T_g(z)$ gradient in our experiment, but also shed light on the puzzling "onset behavior" observed in experiments, where the temperature dependence of τ of polymer nanofilms qualitatively changes just below $T_g^{ne}$. Our findings suggest the presence of new nonequilibrium physics that transforms the magnitude and basic form of near-interface alterations in dynamics and $T_g$ relative to the purely equilibrium limit. This provides valuable insights into the dynamics of nanoconfined glasses and highlights the importance of considering nonequilibrium effects in future studies.

## Results

To measure the $T_g^{ne}(z)$ profiles in cyclic PMMA (*c*-PMMA$_m$) and linear PMMA (*l*-PMMA$_m$) with varying degrees of polymerization (*m*), we have used a method we developed recently (*22, 23*). This method involves labeling the polymer with a fluoro-tracer and determining the temperature at which it gains mobility on heating, which is used as the $T_g^{ne}$. By using a combination of water contact angle measurements, sum-frequency generation vibrational spectroscopy, and X-ray photoelectron spectroscopy to finely adjust the measurement probe depth, *z*, we obtain the local $T_g^{ne}$ at varying *z* with nanoscale resolution.

In Fig. *1*, we present the $T_g^{ne}(z)$ data for the studied *c*-PMMA$_m$. The results indicate that beyond a depth of 1–2 nm from the surface, $T_g^{ne}(z)$ increases linearly with *z* until it reaches the bulk $T_{g,bulk}^{ne}$. The penetration depth, $h_t$ (= 7.8 nm) at which $T_{g,bulk}^{ne}$ is recovered is independent of molecular weight, consistent with previous findings on the near-surface relaxation time of linear polystyrene (*l*-PS) (*24*). The observed linear gradient is also consistent with our *l*-PMMA data (see fig. S1) and published data for *l*-PMMA (*22*) and *l*-PS (*23*). Our findings demonstrate that the experimentally measured $T_g^{ne}(z)$ differs from the exponential form observed for $T_g^{eq}(z)$ in simulations (*15*).

To determine whether this difference arises from nonequilibrium effects, we conduct simulations of glass formation in freestanding long-chain polymer films and their bulk counterpart upon incorporating both extrinsic and intrinsic nonequilibrium effects. To incorporate *extrinsic* nonequilibrium effects, we perform melt-quench simulations in which long-chain conformational statistics begin and remain modestly out of equilibrium, which qualitatively mimics the effect of trapped nonequilibrium chain configurations commonly produced during fabrication of polymer films. To investigate *intrinsic* nonequilibrium effects at simulation versus experimental timescales, we devise a simulation protocol to analyze changes in $T_g^{ne}(z)$ as the simulation time scale ($t_{ANN,max}$, the maximum annealing time applied to the system) increases. By comparing these results with $T_g^{eq}(z)$, we evaluate if nonequilibrium effects can explain the experimentally observed linear $T_g^{ne}(z)$.

To control the nonequilibrium effects, we employ a two-step protocol in our simulations. First, we subject the system to a high-temperature configurational annealing, where we control the extrinsic nonequilibrium effects. Then we thermally quench it to various measurement temperatures ($T$) and perform an isothermal (segmental) annealing for a maximum duration of $t_{ANN,max}$ to control the intrinsic nonequilibrium effects (see Supplementary Simulation Method). Consequently, the system falls out of equilibrium when $\tau_{bulk}$—the relaxation time in the middle of the film, which resembles bulk behavior—exceeds $t_{ANN,max}$. Therefore, $T_{g,bulk}^{ne}$ is approximately the temperature at which $\tau_{bulk} = t_{ANN,max}$. To maintain consistency, we define $T_{g,bulk}^{eq}$ as the temperature at which $\tau_{bulk}$ equals $t_{ANN,max}$ under equilibrium. In our simulation, $t_{ANN,max}$ is initially set to $10^6$ Lennard-Jones time ($\tau_{LJ}$). We measure $\tau$ at $z$ from the intermediate scattering function of the polymers at $z$ (Supplementary Simulation Method). Fig. 2A displays the $\tau$ data obtained at several $z$'s near the surface and in the middle of the film plotted versus $1/T$. Solid lines are fits of the equilibrium data, employing the vertical dotted line as a cutoff (which warrants that $\tau \ll t_{ANN,max}$) to a validated functional form for the temperature dependence of equilibrium dynamics (*25*). At lower temperatures, we continue to employ these fit curves (colored solid curves) to extrapolate the equilibrium dynamics at all the depths examined, as validated previously (*26*). Inside the mid-film, which emulates bulk behavior, the data (orange squares) closely obey the equilibrium line down to temperatures where $\tau$ is greater than $t_{ANN,max}$, showing that $T_{g,bulk}^{ne} \cong T_{g,bulk}^{eq}$, as found before (*20*). In the near-surface region ($z \leq 3.0625$), however, $\tau$ increasingly deviates from the equilibrium curves with decreasing $z$ for $T < T_{g,bulk}^{ne}$, suggesting that intrinsic nonequilibrium effects are more effective in altering dynamics closer to the surface.

Similar to $T_{g,bulk}^{ne}$ and $T_{g,bulk}^{eq}$, we determine $T_g^{eq}(z)$ and $T_g^{ne}(z)$ by finding the temperatures where the equilibrium curves and nonequilibrium data, respectively, intersect the dotted line representing $\tau = 10^6 \tau_{LJ}$ (= $t_{ANN,max}$) in Fig. 2A. The fractional change in $T_g^{ne}(z)$ and $T_g^{eq}(z)$ relative to $T_{g,bulk}^{eq}$ is compared in Fig. 2B. As one can see, $T_g^{eq}(z)$ (filled squares) exhibits an exponential gradient, while $T_g^{ne}(z)$ (orange circles) exhibits a linear regime at the surface. When we decrease $t_{ANN,max}$ to $10^5$, which is equivalent to decreasing the degree of supercooling or increasing the effective cooling rate $q$ (= $1/t_{ANN,max}$), the linear regime of the $T_g^{ne}(z)$ gradient (green diamonds) penetrates less far into the films, while $T_g^{eq}(z)$ (open squares) remains unchanged. In this simulation, $\tau = 10^6 \tau_{LJ}$ corresponds to ~1 μs and the length scale of the bead-spring potential, $\sigma \approx 1$ nm. Since the penetration depth of $T_g^{ne}(z)$ grows with supercooling, a modest naïve extrapolation suggests that the entire gradient would plausibly be linear by the experimental timescale at $T_g^{ne}$, which is consistent with our experimental results.

Our findings above provide insights into the source of the nonequilibrium effects. First, they cannot arise purely from *extrinsic* nonequilibrium (chain conformation) effects because nonequilibrium chain conformations exist above $T_{g,bulk}$, and yet our $T_g^{eq}(z)$ data continue to report an exponential gradient like that in prior simulations of short chains possessing conformational equilibrium (*15*). We further show in figs. S2 and S3 that improving chain equilibrium weakens but does not eliminate deviations of $\tau$ from the equilibrium curves in the vicinity of $T_g^{ne}$. Therefore, it appears that the nonequilibrium effects found here are causally driven, to leading order, by the intrinsic loss of equilibrium near $T_g^{ne}$.

We have observed that the "onset behavior", a long-standing anomaly of nanoconfined glasses, is also found near $T_g^{ne}$. In this behavior, $\tau$ begins to exhibit an Arrhenius $T$-dependence at low temperatures (as opposed to the Vogel-Fucher-Tammann $T$-dependence commonly exhibited by glasses), which has been previously observed (*7, 27, 28*). We explore whether the same *intrinsic* nonequilibrium effects can explain this onset behavior. In Fig. 3A, we demonstrate the onset behavior in our experimental $l$-PMMA$_{173}$ system through the data of $\log(q)$ versus $1/T_g^{ne}$ of nano-films determined by ellipsometry (various symbols). Here, $T_g^{ne}$ denotes the mean-film $T_g$ measured on heating the films at rate $q$. The data shows that the various film thicknesses exhibit Arrhenius $T$-dependences that extrapolate upward to a point (open circle) on or near the bulk curve (solid line). To gain insight into this observation from our simulations, we extrapolate the simulated $T_g^{eq}(z)$ and $T_g^{ne}(z)$ gradients to longer timescales and average them over supported films (on a dynamically non-perturbing substrate) of comparable thicknesses to yield the mean-film values of $T_g^{eq}$ and $T_g^{ne}$ (see Supplementary Simulation Method). The result, shown in Fig. 3B, reveals that the logarithmic rate vs $1/T_g^{ne}$ data (dashed curves) exhibits an upward concavity that is absent in the equilibrium $T_g^{eq}$ data (dotted curves). Mathematically, this is a qualitative consequence of the growing range of the linear nonequilibrium regime with increasing $t_{ANN,max}$ (or $1/q$) (Fig. 2B). Practically, this leads to the emergence at low temperature of a regime of nearly Arrhenius dynamics. If we infer data points from these extrapolations over a timescale range and resolution comparable to various cooling-rate-$T_g$ experiments (shown as solid circles in Fig. 3B), the results essentially replicate the onset behavior reported in Fig. 3A. Specifically, Arrhenius fits to these data (colored solid lines) extrapolated upward towards a convergent point close to the equilibrium bulk curve (black solid line). In contrast, the $T_g^{eq}$ (dotted) curves do not exhibit this convergence, but is projected to continue following the fractional power law decoupling behavior (*29*) previously reported in simulation and predicted in theory. While these curves find an appreciable departure from the bulk curve, it is much weaker than that observed for the nonequilibrium curves. These observations show that the low temperature experimental onset behavior results from a linear gradient in $T_g^{ne}$ caused by intrinsic nonequilibrium effects. The fact that the onset behavior has been observed in small molecule glasses (*28*), which lack the long-ranged conformational modes present in polymers (*27*), indicates that extrinsic nonequilibrium effects are insignificant in the onset behavior. Furthermore, this behavior is found whether $T_g^{ne}$ is obtained by heating (Fig. 3A) or cooling (*27, 28*), which further reinforces the interpretation that intrinsic nonequilibrium effects are the main contributing factor to the onset behavior.

**Discussion and Conclusion**

Our observations suggest a unified picture of altered dynamics and glass formation near a free surface. In equilibrium, surfaces induce an exponential $T_g^{eq}$ gradient along with a weak long-ranged tail, which is consistent with the simulation literature (*15*) and experiments(*17, 18*). However, as the system falls out of equilibrium on cooling through $T_g^{ne}$, intrinsic nonequilibrium effects qualitatively modify and enhance the surface-induced gradient in $T_g$. A linear nonequilibrium gradient progressively grows inward from the surface with increasing equilibration timescale (or equivalently decreasing thermal quench rate), leading to an onset of much stronger and more Arrhenius ($\log q$ vs $1/T_g$) average-$T_g$ shifts. It is this average-$T_g$ shift that is probed in many experiments that report the onset of a new regime of behavior at low temperatures and rates. Besides this onset behavior, several other anomalous experimental observations, such as extraordinarily long-ranged $T_g$ gradients at polymer-polymer interfaces (*11*) and reports of two-distinct $T_g$'s in some thin films systems(*30*), also emanate from $T_g^{ne}$ measurements that we now know are susceptible to nonequilibrium phenomena. This may suggest a new foundation for understanding these remarkable observations by adding nonequilibrium effects into consideration. Results of this work and two recent experiments (*17,*

*18*) suggest that the interfacial dynamical behavior observed in previous computer simulations (*9, 13-15*) likely represents the genuine equilibrium dynamical situation extended to experimental timescales and can be used to validate extant theories of glass transition (*15*) and shed light into the nature of the glass transition.

However, the mechanism by which intrinsic nonequilibrium effects produce deviations of $\tau^{ne}$ from the equilibrium counterpart (Fig. 2A) is unintuitive. Specifically, the deviations are more pronounced closer to the surface even though the relaxation time in that region is shorter and equilibrium can be attained more rapidly. We suggest that intrinsic nonequilibrium effects may be driven by a non-local effect, which originates from a nonequilibrium density profile ($\rho(z)$) that emerges near the surface below $T_{g,bulk}^{ne}$ (*31*). Using simulations, we have obtained one such profile and presented it in Fig. 4A. This profile displays a peak near the surface (at $z = 0$) when the temperature falls below $T_{g,bulk}^{ne}$ (the density data as a function of temperature can be found in table S1). The underlying cause of this phenomenon can be seen in the temperature dependence of the equilibrium (liquid) and nonequilibrium (glassy) densities shown in Fig. 4B. When the temperature falls below $T_{g,bulk}^{ne}$, the mid-film that resembles the bulk starts to follow the weaker temperature dependence of glass density (blue dashed line), reflecting a nonequilibrium decrease in density due to the vitrification of this bulk-like part of the film. As cooling continues below the bulk $T_g$, the density peak near the surface becomes more prominent.

This nonequilibrium effect cannot impact the mid-film $T_g$, because it emerges below the mid-film $T_g$ (upon cooling) as a result of the loss of equilibrium at $T_{g,bulk}$. However, as shown in Fig. 4A, at their local $T_g^{eq}$, molecules just beneath the surface are sandwiched between the surface and a counterintuitively less dense and thus softer vitreous underlayer. Prior theory (*32*), experiment (*11, 33*), and simulation work (*34*) have indicated that nanoscale proximity to a soft domain tends to enhance dynamics. Therefore, in addition to the equilibrium effect of a dynamical enhancement by the free surface, these molecules also experience a nonequilibrium enhancement of dynamics by their proximity to a softer underlayer. Consequently, $T_g^{eq}$ is further suppressed beyond what would be expected from equilibrium dynamics. Moreover, since the relative softness of the underlayer driving this effect increases as the system is cooled more (see Fig. 4B), this additional dynamical enhancement is more pronounced for material closer to the surface, which has a lower $T_g$ even under equilibrium conditions. This explains why the nonequilibrium enhancement in dynamics is most prominent near the free surface as shown in Fig. 2. In essence, this nonequilibrium effect amplifies the equilibrium dynamical gradient with an additional dynamic-enhancing mechanism. However, because this mechanism is based on the underlying equilibrium dynamical enhancement present at the free surface, it would not occur without the underlying equilibrium effect in the first place.

Collectively, our results indicate that nonequilibrium effects play a central role in mediating the thermal and dynamical properties of thin films and nanostructured glasses. Furthermore, those properties can be modulated both by tuning the thermal pathway (and hence $T_g^{ne}$) or trapping nonequilibrium molecular conformations at high temperatures. These possibilities allow for new strategies to rationally control the properties of diverse nanostructured materials. In contrast, most current theories of altered dynamics near interfaces focus on purely equilibrium systems. Given the envisaged opportunities, extending these theories to account for nonequilibrium effects should be a priority in future work.

**Acknowledgments**

JX, XW, OKCT, DSS, and AG would like to express their gratitude for the funding support received for their research. JX and XW were supported by the National Natural Science Foundation of China (grants 22161160317, 22203075, and 22173081). OKCT received support from the Research Grants Council of Hong Kong (grant N_HKUST623/21). DSS and AG were supported by the United States National Science Foundation CBET Program (grant 2208238).

**Supplementary Materials**

Materials and Methods

Figs. S1 to S8

Tables S1 to S3

References (*1–60*)

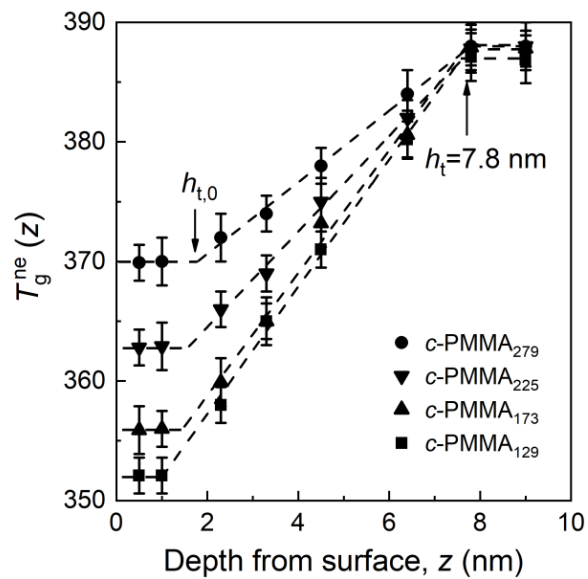

**Fig. 1. Experimental depth profiles $T_g^{ne}(z)$ taken at the surface of various $c$-PMMA polymers.** Dashed lines are the best fits to a linear gradient with details of the fits reported in table S2.

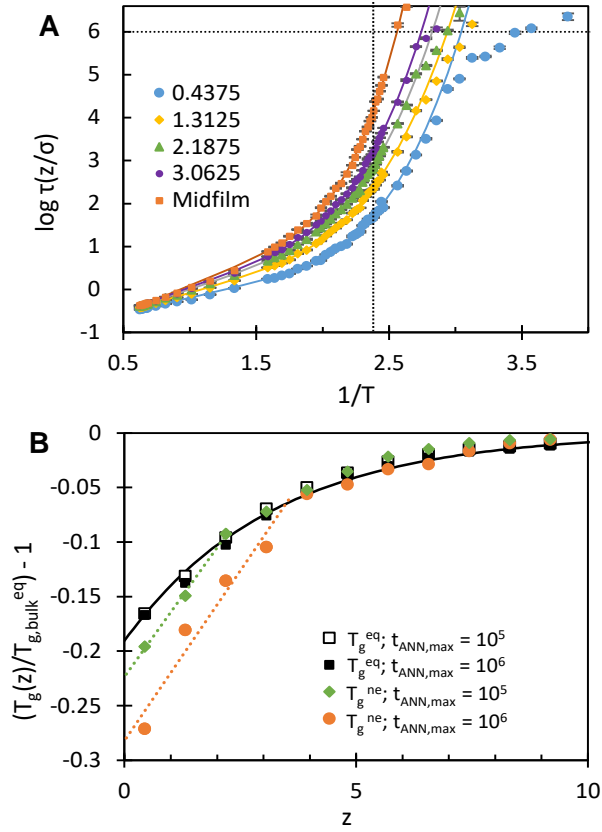

**Fig. 2. Simulation results revealing intrinsic nonequilibrium effects on near-surface τ(z) of long-chain films.** (**A**) Segmental relaxation time τ vs. inverse temperature for various segmental layers at different distances z/σ from the surface (σ is the length scale of the bead-spring potential). The dotted lines represent the low-temperature cutoff for inclusion of equilibrium relaxation time data and $\tau = t_{ANN,max} = 10^6$. The curved lines show fits of the data to an established model for equilibrium dynamics (*26*). Deviations from the curved equilibrium lines to the right of the vertical dotted line indicate intrinsic nonequilibrium effects on the near-surface relaxation time near and below $T_{g,bulk}^{eq}$. (**B**) Normalized alteration gradients of $T_g^{eq}$ and $T_g^{ne}$ relative to bulk for systems with $t_{ANN,max} = 10^5$ and $10^6$ (inicated in the legend). The solid line show a fit of $T_g^{eq}$ (z) to an exponential gradient form to the $T_g^{eq}$ data, while the dotted lines (in corresponding colors) represent linear fits to the surface region where $T_g^{ne}$ deviates significantly from $T_g^{eq}$.

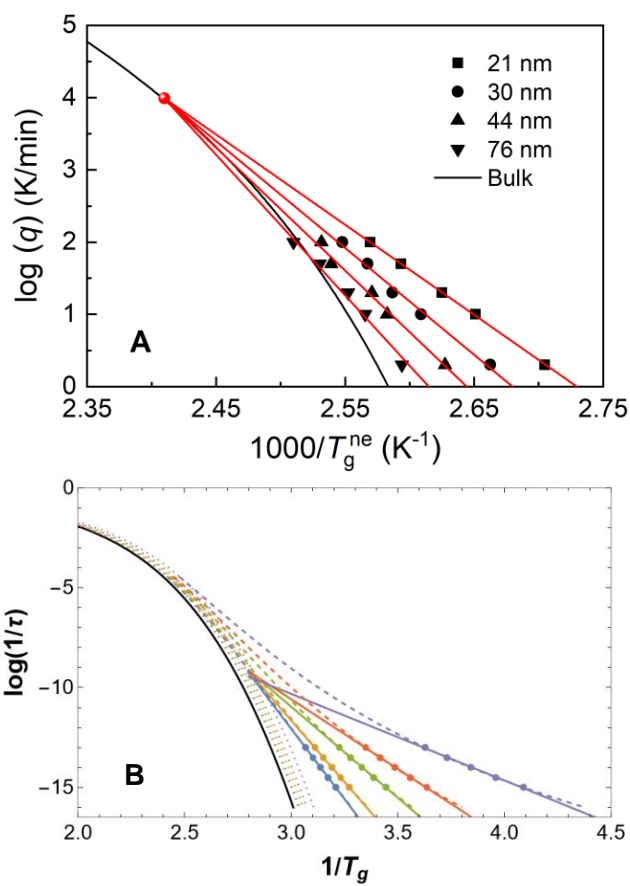

**Fig. 3. Experimental and simulation data showing the onset behavior of $\log(1/\tau)$ vs. $1/T_g^{ne}$ in polymer nanofilms** (**A**) Experimental analysis of mean-film $T_g^{ne}$ for $l$-PMMA$_{173}$ films of varying thicknesses (indicated in the legend) and heating rates ($q$). The red lines represent fits to Arrhenius rate laws, extrapolating upwards to an intercept (open circle) near the equilibrium bulk curve (solid black line), similar to prior $T_g^{ne}$ measurements conducted during cooling. (**B**) Simulated results on mean logarithmic relaxation rate $\log(1/\tau)$ for a supported film on a dynamically neutral substrate plotted vs. $1/T_g$, for comparison to experimental data. (See supplementary Simulation Method for details) Dashed lines show predicted behavior for mean-film $T_g^{ne}$ with different thicknesses ($h$ = 20 σ, 30 σ, 40 σ, 60 σ, and 75 σ—top to bottom), while dotted lines show predicted behavior of $T_g^{eq}$ for the same systems. Points represent "pseudodata" mimicking $T_g^{eq}$ data that could be obtained in experiment, along with solid lines passing through them representing Arrhenius fits. These lines extrapolate upwards to a point near the equilibrium bulk curve.

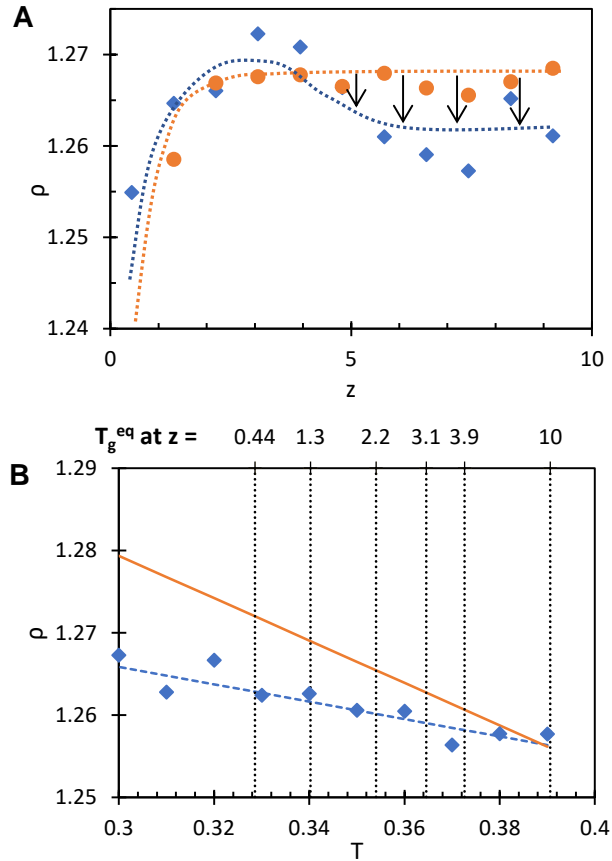

**Fig. 4. Illustrating the proposed relationship between nonequilibrium density gradient effects and nonequilibrium $T_g$ gradient effects.** (**A**) Nonequilibrium glassy density (blue diamonds) and equilibrium liquid density extrapolated from high-temperature melt data (orange circles) vs. distance (in units of $\sigma$) from the film surface at a simulation temperature of ~0.35, below the system's bulk $T_g$ (which is ~0.4). Dotted lines are guides-to-the-eye. The nonequilibrium suppression in $T_g$ in the vitreous interior is posited to serve as a soft, $T_g$-reducing underlayer at lower temperatures, where the near-surface material in the local $T_g$ maximum vitrifies. (**B**) Temperature dependence of the extrapolated equilibrium density (orange line) and of the nonequilibrium, glassy state density of the film interior below the bulk $T_g$ (blue diamonds and dashed guide-to-the-eye line). The vertical dotted lines correspond to nonequilibrium local $T_g$ values at distances from the surface, as labeled at the top of the panel. The length of the arrows in panel A, showing nonequilibrium suppression in density in the film interior, corresponds to the difference between the two lines in panel B. This difference grows on cooling, resulting in a larger difference at the local $T_g$ nearer the surface. Consequently, a relatively softer (and more $T_g$-suppressing) underlayer material is present nearer the surface.

# Supplementary Materials for

## Mixed equilibrium/nonequilibrium effects govern surface mobility in polymer glasses

Jianquan Xu[1], Asieh Ghanekarade[2], Li Li[1], Huifeng Zhu[1], Hailin Yuan[3], Jinsong Yan[3], David S. Simmons[2]\*, Ophelia K. C. Tsui[3,4]\*, Xinping Wang[1]\*

Corresponding authors: David S. Simmons, dssimmons@usf.edu; Ophelia K. C. Tsui, okctsui@ust.hk; Xinping Wang, wxinping@zstu.edu.cn

**The PDF file includes:**
    Materials and Methods
    Figs. S1 to S8
    Tables S1 to S3
    References (*1–60*)

# 1. Materials

Methyl methacrylate (MMA, 99%), toluene (anhydrous, 99.8%), 2-bromoisobutyryl bromide (BIBB, 98%), N,N,N',N'',N''-pentamethyl diethylenetriamine (PMDETA, 99%) and triethylamine (TEA, 99%) were purchased from Sigma-Aldrich Co., USA. 2-perfluorooctylethyl methacrylate (FMA, 97%) and Benzotrifluoride (BTF, 99%) were purchased from J&K Scientific Ltd. 3-(trimethylsilyl)-2-propyn-1-ol (TMS-C≡C-CH$_2$OH, 98%), Copper(I) bromide (CuBr, 99%), Copper(II) bromide (CuBr$_2$, 99%) and tetrabutylammonium fluoride (TBAF, 1.0 M in tetrahydrofuran) were acquired from Aladdin Chemical Co., Ltd. Prior to polymerization, MMA and FMA were purified by passing them through a basic alumina column (100-200 mesh) and dried over CaH$_2$. BTF was also dried over CaH$_2$ and then distilled under reduced pressure. CuBr was purified by an acetic acid (aqueous) solution. All the other reagents and solvents, such as tetrahydrofuran (THF), dimethylformamide (DMF) and sodium azide were analytical grade and used as received.

1.1. Synthesis of the linear and cyclic PMMA with a fluorinated group label

Linear (*l*-) and cyclic (*c*-) PMMA labeled by the fluoro-group FMA were synthesized by combining atom transfer radical polymerization (ATRP) (*35*) and copper (I)-catalyzed alkyne-azide cycloaddition (CuAAC) "click" reaction (*36, 37*). Figure S4 shows the detailed synthetic route, where the serial numbers **1**-**6** denote the different intermediate chemical products produced at different stages of the route as discussed below.

  The alkyne functionalized initiator, TMS-C≡C-CH$_2$OOCC(CH$_3$)$_2$Br (**2**), was synthesized as published in reference (*38*). A 50 mL of THF solution with BIBB (5.6 ml, 50.6 mmol) was added dropwise to a solution of TMS-C≡C-CH$_2$OH (**1**) (5 ml, 33.7 mmol) and TEA (7ml, 50.6 mmol) in 90 ml THF, while maintaining the reaction temperature at 273 K. After all the solution had dripped off, the reaction solution was stirred for another 5 h at room temperature for the reaction to complete. The triethylammonium salt was removed from the reaction solution by filtration then the solvent was removed by using a rotary evaporator. Afterwards, the crude product was dissolved in CH$_2$Cl$_2$ and washed twice with saturated solution of NH$_4$Cl and twice more with distilled water. The resulting organic layer was dried with anhydrous MgSO$_4$ and subsequently purified by column chromatography (petroleum ether/ EtOAc 95:5). The final product looks like a colorless oil and confirmed by $^1$H NMR spectra. Owing to the protection of the neighboring TMS group, complexation of the alkyne group in the initiator by the copper catalyst and side reactions such as Glaser coupling (*39*) were circumvented.

  The macroinitiator *l*-TMS-C≡C-PMMA$_m$-Br (**3**) and fluoro-tracer-labeled linear PMMA (*l*-TMS-C≡C-PMMA$_m$-b-FMA$_n$-Br (**4**)) with different molecular weights were synthesized by ATRP similar to our previous work (*35*). But here, we used Cu (II) as catalyst (which was reduced by MMA to Cu(I)) rather than Cu(I) directly (which is insensitive to moisture and oxygen) to yield lower molecular weights and narrower molecular weight distributions. The number of FMA groups introduced to each chain is 1~2, as confirmed by $^{19}$F NMR using an internal standard method with trifluorotoluene as the internal standard substance. After obtaining *l*-TMS-C≡C-PMMA$_m$-*b*-FMA$_n$-Br (**4**), azide reaction was executed (*37*). Three grams of *l*-TMS-C≡C-PMMA$_m$-*b*-FMA$_n$-

Br (**4**) was dissolved in 10 ml of DMF in a 50 ml three necked flask equipped with a magnetic stirrer. NaN$_3$ (10 equiv.) was added to this solution and the mixture was stirred for 24 h at 298 K. Then the polymer was precipitated in cryogenic methanol. Thereafter, 2g of *l*-TMS-C≡C-PMMA$_m$-b-FMA$_n$-N$_3$ (**5**) was dissolved in 10 ml of THF in a 50 ml three necked flask with 1.5 equiv. of TBAF and stirred for 24 h at 298 K (*40*). Then *l*-C≡C-PMMA$_m$-b-FMA$_n$-N$_3$ (**5**) was precipitated in cryogenic methanol. The cyclic PMMA was synthesized by copper (I)-catalyzed alkyne-azide cycloaddition (CuAAC) "click" reaction (*36, 37*). A 0.2 mM solution of *l*-C≡C-PMMA$_m$-b-FMA$_n$-N$_3$ (**5**) in toluene (100 mL) was added in three freeze / pump / thaw cycles in a 250ml flask. In another 500 ml flask, PMDETA (100 equiv.) was dissolved in toluene (200 mL) and degassed in three freeze / pump / thaw cycles, then Cu(I)Br (50 equiv.) was added immediately, followed by two pump-and-thaw cycles. Upon thawing, a syringe with pump was employed to transfer the solution of *l*-C≡C-PMMA$_m$-b-FMA$_n$-N$_3$ (**5**) to a rapidly stirring solution of Cu(I)Br/PMDETA in toluene at room temperature at a rate of 8 µL/min. After the addition of the polymer solution was completed, an additional 2 h was allowed to stir the solution. The final product, *c*-PMMA$_m$-b-FMA$_n$ (**6**), was purified by neutral aluminum oxide to remove the copper salt, followed by precipitation in n-hexane and drying in vacuum. FTIR, $^1$H NMR, and GPC were employed to check that the "click" reaction occurred successfully (figs. S5A and S5B) and the product was *c*-PMMA$_m$-b-FMA$_n$ (fig. S5C).

1.2. Film preparation

Silicon (100) wafers covered by a native oxide layer (SiO$_x$, ~2 nm) were cut into 1.3 cm × 1.3 cm slides and employed as a substrate. The slides were cleaned by a freshly prepared piranha solution (H$_2$SO$_4$:H$_2$O$_2$ (v:v) = 3:1) at 363 K for 30 min. The cleaned substrates were coated by gold by sputtering in a sputter Q 150T S (Quorum, U.K.) under an argon atmosphere. The thickness of the gold layer was about 140 nm, as determined by atomic force microscopy (AFM). Thin films of *c*-PMMA$_m$-b-FMA$_n$ films with different thicknesses were obtained by spin-coating solutions of the polymers in cyclohexanone at various concentrations onto the gold-coated substrates. Afterwards, the films were dried in vacuum at 318 K for 72 h to remove the residual solvent. The film thickness of was determined by an EP$^3$SW ellipsometer (Accurion GmbH Co., Germany). Before taking any measurements, the films were annealed for a period of 24 h at each measurement temperature.

1.3. Sample characterization

The molecular weight distributions of the polymers were determined by GPC (Waters-150C, USA). Solutions of the polymers in THF at a concentration of 1 mg/ml were injected into the GPC at 313 K at a flow rate of 1 mL/min. Differential Scanning Calorimetry (DSC) (Q2000, TA Instruments, USA) under a nitrogen environment was used to measure the $T_g$ ($T_{g,\text{bulk}}^{\text{ne}}$ (DSC)) of the *c*-PMMA bulk polymers. During measurement, the sample was first heated to 453 K and held for 5 minutes to eliminate thermal history. Then it was cooled to 323 K at the fastest rate and kept at 323 K for 1 min. Afterwards, the sample was heated at 2 K/min to 453 K during which $T_{g,\text{bulk}}^{\text{ne}}$(DSC) was obtained. The radii of gyration ($R_g$) of the *c*- and *l*-PMMA polymers were calculated by $R_g = (Nl_b^2/12)^{1/2}$ and $R_g = (Nl_b^2/6)^{1/2}$, respectively (*41*) (where $l_b$ is the average statistical segment length, which is 0.69 nm for PMMA, and $N$ is the degree of polymerization).

## 2. **Methods**

2.1. Experimental Methods

### 2.1.1. *Method for measuring depth profile of $T_g^{ne}(z)$*

We employed a method that we recently developed (*22, 23*) to measure the depth $z$ profile of $T_g^{ne}$ for fluoro-labeled *c*-PMMA$_m$-b-FMA$_n$ and *l*-PMMA$_{173}$-b-FMA$_1$ polymers with varying $m$ and corresponding number-average molecular weights, $M_n$ = 12.9–27.9 kg/mol. This method involves three surface-sensitive techniques—water contact angle measurement (WCA), sum-frequency generation vibrational spectroscopy (SFG), and angle-resolved X-ray photoelectron spectroscopy (XPS)—which have different probe depths of 0.5 nm (*42*), 1 nm (*35*), and 2.3–9.0 nm (*22, 23*), respectively.

Because the fluoro-labels have a lower surface energy than PMMA, they tend to migrate to the free surface. At a given temperature $T$ between the $T_g^{ne}$ at $z = 0$ and the bulk counterpart $T_{g,bulk}^{ne}$, this migration process is halted for $z > z^*$, where $z^*$ is the position for which $T_g^{ne}(z^*) = T$. This occurs simply because the local $T_g^{ne}(z)$ deeper in the film than $z^*$ are greater than the experimental temperature, and so the segmental dynamics (which are the main dynamics that govern the glass transition in polymers) are frozen in the glassy state. Therefore, the temperature at which the fluoro labels within depth $z$ begins to enrich on heating (because of migration of the fluoro labels from the region immediately below) is taken to be the local $T_g^{ne}$ at $z$. By exploiting the fact that the different surface techniques have different probe depths, $T_g^{ne}(z)$ could be measured from $z = 0.5$–9.0 nm. In all our measurements, a 24-h waiting time was allowed upon each stepwise increase of $T$ for the sample to stabilize. The resulting $T_g^{ne}$ determination is akin to a devitrification temperature measurement because it probes the temperature at which degrees of freedom at a given depth z are liberated on heating.

Figure S6A shows a schematic of the experimental method. Fig. S6B displays representative data from WCA measurement and SFG ($I_{2910}/I_{2950}$) plotted as a function of $T$. As we can see, the WCA (solid squares) stays constant at low $T$ then begins to increase with $T$, indicating an increase in the hydrophobicity of the film surface when $T$ exceeds 356.0 K. This is understood to be caused by migration of the fluoro groups to the free surface when the lowest near-surface $T_g$ (namely, $T_g^{ne}(z = 0)$) was surpassed (*23, 43*). Because the probe depth of the WCA measurement is 0.5 nm, we assigned $T_g^{ne}(z = 0.5$ nm$)$ to be 356.0 K. For the SFG measurement, we assessed the local glass transition by using the ratio of the SFG peak heights at wave vectors 2910 and 2950 cm$^{-1}$ ($I_{2910}/I_{2950}$), which originate from the C-H symmetric stretching vibrations of the -CH$_2$- groups in the backbone and the ester methyl (-OCH$_3$) groups in the side chains of PMMA, respectively (*44*). The data (open circles) shows that $I_{2910}/I_{2950}$ is constant, equal to the maximum value at low $T$, but begins to decrease with $T$ from 355.9 K (fig. S6B). The steady decrease seen in $I_{2910}/I_{2950}$ with $T$ indicates that the polymer backbones have rearranged and so we infer that devitrification has occurred. Because the probe depth of SFG is 1.0 nm, we assigned $T_g^{ne}(z = 1.0$ nm$)$ to be 355.9 K. The XPS data, represented by the ratio of the F and C content (F/C), are shown in fig. S6C. In all these data, F/C is a constant at low $T$, but begins to increase with $T$ at an onset temperature that increases with the XPS probe depth ($z = 3\lambda \sin\theta$, where $\theta$ is the take-off angle measured from the sample surface as illustrated in fig. S6A, and $\lambda = 3$ nm is the mean free path for inelastic scattering in PMMA with an Al Ka X-ray source of energy 1253.6 eV(*45*)). We similarly assigned the onset temperature for F/C to increase to be the local $T_g^{ne}$ at $z$.

To carry out WCA measurements, we used a Krüss DSA-10 CA goniometer (Hamburg, Germany). In each measurement, a ~2 μl drop of deionized water was dispensed onto the sample surface. After a waiting time of 3 s, the contact angle of the droplet was measured.

Sum frequency generation (SFG) spectra were collected by using a custom designed SFG spectrometer (EKSPLA, Lithuania). During measurement, the incident angles of a tunable infrared (IR) laser beam and a 532 nm visible laser beam were fixed at 50° and 70°, respectively, to generate sum frequency (SF) signal in a second-order nonlinear optical process. The ssp (SF/*s*, visible/*s*, and IR/*p*) mode and IR/p polarization sum-frequency signals were collected at a frequency of 4 cm$^{-1}$, where *s* (*p*) represents the polarization normal (parallel) to the plane of incidence. Fig. S7 displays a typical series of *ssp* SFG spectra taken of a *c*-PMMA$_{173}$-b-FMA$_1$ film surface at various $T$ across the $T_g^{ne}$ (= 355.9 K) within the probe depth of SFG. It shows that the peak intensity at 2910 cm$^{-1}$ ($I_{2910}$), originated from C-H symmetric stretching vibrations of the -CH$_2$- groups in the backbone, gradually decreased with $T$ from 353 to 393 K. At the same time, the peak intensity at 2950 cm$^{-1}$ ($I_{2950}$), originated from the -OCH$_3$ groups in the side chains of PMMA (*44*), was relatively unaffected. Therefore, the steady decrease seen in $I_{2910}$ (relative to $I_{2950}$) with $T$ above ~353 K indicates that the polymer backbones have rearranged, and so kinetic freezing has occurred.

Angle-resolved XPS spectra were conducted by using a PHI5000C ESCA system with an Al Kα X-ray source (1253.6 eV) operated at 250 W and 140 kV. The analytical depth ($z$) of the XPS analysis was estimated by $z = 3\lambda \sin\theta$ (*46*), where $\theta$ is the emission angle of the photoelectron to the surface plane and $\lambda$ is the mean-free-path for inelastic scattering, which is ~3 nm for PMMA(*45*), allowing for varying probe depths of ~2.3–9.0 nm.

Next, we analyze the $T_g^{ne}(z)$ data of the *c*-PMMA$_m$-b-FMA$_n$ films, shown in Fig. *1*. It is noted that $T_g^{ne}(z)$ remains constant for $z < 1.0$–2.0 nm. Beyond this depth, $T_g^{ne}(z)$ increases linearly with $z$ until it reaches the bulk value at $T_{g,bulk}^{ne}$. The profile of $T_g^{ne}(z)$ can be mathematically described by

$$T_g^{ne}(z) = \begin{cases} T_g^{ne}(0) & z \leq h_{t,0} \\ T_g^{ne}(0) + \frac{T_{g,bulk}^{ne} - T_g^{ne}(0)}{h_t - h_{t,0}}(z - h_{t,0}) & h_{t,0} < z \leq h_t \\ T_{g,bulk}^{ne} & z > h_t \end{cases} \quad (S1)$$

where $h_{t,0}$ and $h_t$ is the depth where the linear dependence of $T_g^{ne}(z)$ begins and ends, respectively (Fig. *1*). By fitting equation S1 to the $T_g^{ne}(z)$ data in Fig. *1*, and using the value of $T_{g,bulk}^{ne}$ determined by ellipsometry (as described in the next section), we infer the values of $T_g^{ne}(0)$, $h_{t,0}$, $h_t$, and $[T_{g,bulk}^{ne} - T_g^{ne}(0)] / [h_t - h_{t,0}]$, which are presented in table S2.

We have also conducted similar measurements on *l*-PMMA films with various degrees of polymerization. The $T_g^{ne}(z)$ gradients in these films also show a linear increase with $z$. However, unlike $T_g^{ne}(z)$ of the *c*-PMMA films, these data do not reach $T_{g,bulk}^{ne}$, and therefore, the full profile of $T_g^{ne}(z)$ cannot be determined. Nevertheless, our simulation data in Fig. 2B demonstrate that the profile of $T_g^{ne}(z)$ develops a linear portion near the surface, which penetrates further into the film with increasing degree of supercooling. This suggests that the full profile of $T_g^{ne}(z)$ would possibly be linear under experimental conditions.

2.1.2. *Method for measuring the mean-field $T_g^{ne}$ by ellipsometry*

To determine the mean-film $T_g^{ne}$ for films of varying thicknesses, $h$, we conducted experiments using an EP$^3$SW ellipsometer (Accurion GmbH Co., Germany) to measure the ellipsometric angles, Delta and Psi, while increasing the temperature $T$ at a steady rate. Then we fitted the data using the method reported in Ref. (*47*). Similar to the $T_g^{ne}(z)$ study, the films were not subjected to thermal annealing before mean-film $T_g^{ne}$ measurement.

To determine $T_{g,bulk}^{ne}$, we conducted mean-film $T_g^{ne}$ measurements on films with a bulk thickness of 106 nm. To ensure consistency with the $T_g^{ne}(z)$ study, we adopt the same thermal scheme, allowing for a 24-hour waiting time between temperature changes. We found that the $T_{g,bulk}^{ne}$ value determined for the *c*-PMMA$_{173}$ films with a bulk thickness of 106 nm (~387.5 K; fig. S8) was the same as the corresponding value obtained by XPS at $z = 7.8$ and 9.0 nm (~387.8 K; fig. S6C) within experimental error. Furthermore, this value of $T_{g,bulk}^{ne}$ is ~6 K lower than that obtained by DSC (~393.4 K; table S3), attributable to the slower measurement rate used in the $T_g(z)$ measurements compared to the DSC measurements. For other *c*-PMMA's, we found that the relation, $T_{g,bulk}^{ne} = T_{g,bulk}^{ne}$ (DSC) – 6 K also holds for $T_{g,bulk}^{ne}$ determined under the same thermal scheme as in the $T_g^{ne}(z)$ study.

2.2. Simulation Methods

2.2.1. *Model*

Molecular dynamics simulations employ an attractive variant of the bead-spring model of Kremer-Grest (*48*), which has been employed in our prior work probing thin-film dynamics and glass formation behavior in simulation. We primarily focus on chains comprised of 400 beads, although below we also include a result from a 1000-bead chain.

Within this model, non-bonded beads interact via the 12–6 Lennard-Jones potential, with energy and range parameters of $\varepsilon = 1$ and $\sigma = 1$, respectively, and with a cutoff distance of 2.5 $\sigma$. Bonded beads, on the other hand, interact via the Finitely Extensible Nonlinear Elastic Potential,

$$E_b = -\frac{1}{2} K R_0^2 \ln\left[1 - \left(\frac{r}{R_0}\right)^2\right] + 4\varepsilon_b \left[\sigma_b^{12} r^{-12} - \sigma_b^6 r^{-6}\right] \tag{S2}$$

with $K = 30$, $R_0 = 1.3$, $\varepsilon_b = 1.0$, and $\sigma_b = 0.8$. These bond parameters are slightly different from the standard values, and yield a shorter bond length that has been shown to better suppress crystallization without significantly altering segmental dynamics (*49*).

Simulations are performed in the LAMMPS molecular dynamics package (*50*). We simulate the polymers in both a freestanding thin film configuration and a bulk system to compare their behaviors. For the film simulation, we use a nominally NVT ensemble, but due to the presence of free surfaces, the pressure is practically constant at $P = 0$ within the film. For the bulk simulation, we use an NPT ensemble at $P = 0$. To control the pressure within the film, we use the Nose-Hoover thermostat implemented in LAMMPS, while for the bulk simulation, we use the Nose-Hoover thermostat/barostat pair, also implemented in LAMMPS.

2.2.2. *Analysis for estimating the relaxation time $\tau$ from simulation*

Translational relaxation dynamics are quantified via the self-part of the intermediate scattering function, computed at a wavenumber of 7.07, approximately equivalent to the first peak in the segmental structure factor. To extract the relaxation time, $\tau$, from these functions, the long-time portion of the time dependence is fit to a Kohlsrauch-Williams-Watts (KWW) stretched exponential function, which allows us to smooth and interpolate the data. The relaxation time is then determined as the timescale at which this function relaxes to a value of 0.2. This method yields a value similar to the zeroth moment of the long-time $\alpha$ process, over the set of stretching exponents $\beta$ encountered in this polymer. These conventions are commonly used in simulation literature (*51-54*). We perform this analysis at a local level by sorting beads into bins that correspond to a layer thickness $\sigma$, based on their distance from a free surface. From these data, we extract the equilibrium and nonequilibrium glass transition temperatures for each bin.

2.2.3. *Thermal protocols producing extrinsic nonequilibrium effects*

We generate initial randomized configurations using the Packmol package and subject them to a high temperature isothermal annealing procedure at a reduced Lennard-Jones temperature $T = 1.5$ for varying durations of $5 \times 10^5$ $\tau_{LJ}$, $2.5 \times 10^6$ $\tau_{LJ}$, or $5 \times 10^6$ $\tau_{LJ}$. It is worth noting that the initial anneal occurs at a temperature ~4.4 times the experimental $T_g$ of the polymer models in this class, which is around $T = 0.34$ on the 100-second timescale. This initial annealing period serves to alleviate unfavorable chain configurations associated with the initial state, similar to what is observed in many thermally quenched experimental systems. To track the evolution in chain conformational statistics, we compute the mean end-to-end distance of the chains after these distinct annealing periods. The root-mean-square end-to-end distance is computed as 20.1 $\sigma$, 22.5 $\sigma$, and 23.5 $\sigma$ after annealing periods of $5 \times 10^5$ $\tau_{LJ}$, $2.5 \times 10^6$ $\tau_{LJ}$, and $5 \times 10^6$ $\tau_{LJ}$, respectively. These results suggest that chain conformations are modestly removed from equilibrium for the shortest annealing periods and evolve towards it with increasing time. Throughout this paper, we use the term "*extrinsic* nonequilibrium effects" to describe the presence of out-of-equilibrium chain conformations.

In addition to the above protocol, we also present results using 1000-bead polymer chains. In this case, we employ the hybrid molecular dynamics / Monte Carlo chain-rebridging protocol of Sides et al (*55*) during the high-temperature chain configurational annealing procedure. This protocol allows for a much faster equilibration of chain conformational modes. We anneal the system while employing this algorithm for $10^9$ timesteps, which is equivalent to $5 \times 10^6$ $\tau_{LJ}$.

As demonstrated by figs. S2 and S3, the presence of extrinsic nonequilibrium affects the system dynamics only when intrinsic nonequilibrium effects are also present. Furthermore, their degree of influence correlates with the strength of the extrinsic nonequilibrium effects, resulting in a greater deviation of relaxation time from the equilibrium relaxation time ($\tau^{eq}$).

2.2.4. *Thermal protocols to produce systems without intrinsic nonequilibrium effects*

Unlike extrinsic nonequilibrium effects, intrinsic nonequilibrium effects are governed by segmental equilibrium. Our prior work (*56*) showed that subjecting the system to isothermal annealing at a specific temperature $T$ for a duration of 10 times $\tau^{eq}$ is more than enough to establish segmental equilibrium within the system at $T$. Based on this finding, we have devised a thermal protocol to introduce intrinsic nonequilbirum effects to the system. But first, we outline the protocol for creating systems at segmental equilibrium that are free of intrinsic nonequilibrium effects.

Following the initial chain conformational annealing procedure, we subject the systems to thermal quenching from the high temperature using the Predictive Stepwise Quench (PreSQ) algorithm, which we developed earlier (*56*). This algorithm employs an iterative quench-and-anneal process to gather relaxation data as the targeted in-equilibrium relaxation timescale is approached. To achieve segmental equilibrium at a lower temperature $T$, the system undergoes an initial quench from the conformational annealing temperature, with the configurations periodically saved at temperatures along the quench. Subsequently, the configuration at temperature $T$ is subjected to an isothermal anneal for a duration of $10\,\tau^{eq}(T)$. However, because the simulated chains are quite long, these post-quench segmental anneals are generally not long enough to allow for re-equilibration of whole chain conformations. Thus, the whole chain conformational state, and any associated extrinsic nonequilibrium effects, are largely set by the initial high-temperature anneal. Following segmental annealing, dynamical data are collected at each $T$ from a data collection run.

2.2.5. *Protocol to introduce intrinsic nonequilibrium effects to the system and determination of $T_{g,bulk}^{ne}$, $T_{g,bulk}^{eq}$, and $T_g^{ne}(z)$*

In our simulations, we set a maximum limit for the segmental annealing time, denoted by $t_{ANN,max}$. In our primary simulations, this limit is set to $10^6\,\tau_{LJ}$, where $\tau_{LJ}$ is the Lennard-Jones time unit. Above $T^*$—the temperature at which $\tau^{eq}(T) = t_{ANN,max}$—we subject the system to annealing periods of $10\tau^{eq}$, which would allow the system to attain segmental equilibrium. Below $T^*$, the system is permitted to deviate from segmental equilibrium according to the following protocol. Starting from $T^*$, we quench the system to a lower temperature while saving its configurations at regular temperature intervals. Each configuration is then subjected to a *fixed* isothermal annealing period of $t_{ANN,max}$. Because $\tau^{eq}$ exceeds $t_{ANN,max}$ below $T^*$, $T^*$ is assigned to be $T_g^{ne}$. The timescale for the nonequilibrium glass transition to occur is $t_{ANN,max}$ as it is the relaxation time value $\tau^{ne}$ at $T_g^{ne}$.

To compare the equilibrium and nonequilibrium $T_g$'s, we also use $t_{ANN,max}$ as the timescale for the equilibrium glass transition to occur. Using this criterion, we can similarly determine the equilibrium bulk $T_g$ ($T_{g,bulk}^{eq}$). To determine the local $T_g^{ne}$ at a given depth $z$, we assign it as the temperature at which the local segmental relaxation time at $z$ exceeds $t_{ANN,max}$.

### 2.2.6. *Controlling the effective quench rate by adjusting $t_{ANN,max}$*

The effective quench rate for each simulation is controlled by the inverse of the maximum isothermal segmental annealing period ($1/t_{ANN,max}$). Our longest-time simulation employs a maximum segmental annealing time of $10^6$ $\tau_{LJ}$, resulting in the lowest effective quench rate and the deepest supercooled state achieved. To investigate the influence of the depth of supercooling on the $T_g^{ne}$ gradient, we conduct simulations using the PreSQ protocol to maintain equilibrium only down to the temperature at which the bulk relaxation time is $10^4$ or $10^3$ $\tau_{LJ}$. Correspondingly, we lower $t_{ANN,max}$ to $10^5$ or $10^4$ $\tau_{LJ}$, respectively, which is equivalent to increasing the quench rate by a factor of 10 or 100, respectively. These changes lead to a loss of equilibrium at a higher temperature, and an increase in $T_g^{ne}$.

### 2.2.7. *Extrapolation of low-temperature behavior and construction of Fig. 3B in the main text*

To determine the extrapolated low-$T$ simulation behavior shown in Fig. 3*B*, we follow the procedure below. We compute the extrapolated in-equilibrium $\tau^{eq}$ and nonequilibrium $\tau^{ne}$ data at low temperatures for a film supported on a dynamically neutral substrate, which provides a reasonable qualitative basis of comparison with our experiments. To accomplish this, we begin with the observed *equilibrium* gradient of $\tau^{eq}$ at the simulated film surface, which has a double-exponential form:

$$\ln\left(\frac{\tau_{bulk}(T)}{\tau^{eq}(z,T)}\right) = A\exp\left(-\frac{z}{\xi}\right), \quad (S3)$$

with parameters $A = 0.19$ and $\xi = 3.2$, as determined by simulations in the thick film limit. Although our simulations involve freestanding films, we can predict the behavior of supported films by averaging over a single gradient and truncating it at the second surface. This approach was validated in our recent work, which examined equilibrium dynamical gradients in supported films on dynamically neutral substrates (*57*). In this study, we calculate the mean film equilibrium relaxation time $\tau(h,T)$ for a film of thickness $h$ by computing the linear arithmetic average over the film relaxation time gradient as follows:

$$\tau^{eq}(h,T) = \frac{1}{h}\int_0^h \tau^{eq}(z,T)dz, \quad (S4)$$

with $\tau^{eq}(z,T)$ being given by equation S3. Prior studies have shown that the equilibrium gradient's $A$ and $\xi$ values are insensitive to film thickness and temperature (at low simulation temperatures) (*15, 29, 32, 58-60*). Therefore, we can use a combination of equations S3 and S4 to estimate the shift in relaxation time relative to bulk at temperatures beyond those accessed by simulation. The only additional requirement is to extrapolate the bulk relaxation time to lower temperatures. To accomplish this, we employ the Cooperative Model of Schmidtke et al. (*25*). As discussed in the main text, we previously employed experimental data over a broad range of timescales to show that this functional form provides excellent extrapolations of bulk equilibrium relaxation times from simulation timescales to experimental timescales (*26*).

Using the above protocol, we generated the dotted lines for the equilibrium relaxation times in Fig. 3B. To obtain the extrapolated nonequilibrium relaxation times (represented by the dashed lines), we employ a simple extrapolation of the behavior reported in Fig. 2B. In this figure, the $T_g^{ne}$ gradient exhibits a linear regime near the film surface, which recovers the $T_g^{eq}$ gradient at large $z$ and penetrates more deeply into the film with increasing $t_{ANN,max}$. Therefore, we model the nonequilibrium $\tau^{ne}$ gradient as follows:

$$\ln\left(\frac{\tau_{bulk}(T)}{\tau^{ne}(z,T)}\right) = \begin{cases} b - mz & z < z^* \\ A\exp\left(-\frac{z}{\xi}\right) & z > z^*, \end{cases} \tag{S5}$$

where $z^*$ is defined as the intersection point of the low-$z$ linear and high-$z$ exponential regimes of $\ln(\tau_{bulk}(T)/\tau^{eq}(z,T))$. Here, $A$ and $\xi$ employ the values reported above. Fits to the linear regime of $\tau^{ne}(z)$ suggests that $m \cong 0.06$ is essentially temperature-invariant for this system. The growing penetration depth of the linear regime with increasing $t_{ANN,max}$ (or $\tau_{bulk}$) is driven by a growth of the parameter $b$ with increasing supercooling at $T_g^{ne}$. However, with only two data points of $t_{ANN,max}$ (Fig. 2B), we cannot confidently assess the form of this dependence. To provide a leading-order qualitative extrapolation, we treat $b$ as linear in $t_{ANN,max}$ which yields $b = -0.069 + 0.058 \ln(t_{ANN,max})$.

To obtain the extrapolated nonequilibrium behavior, we combine equation S5 with the average provided by equation S4, which yields the dashed lines in Fig. 3B. The upward concavity of these lines is a direct qualitative consequence of the growth of $b$ with increasing $t_{ANN,max}$ (corresponding to decreasing $T_g^{ne}$). While the assumption of a linear dependence of $b$ on $t_{ANN,max}$ is relevant to the precise quantitative shape of this turnover, its qualitative presence is expected to be relatively insensitive to the functional form of this extrapolation.

The black solid line in Fig. 3B is the equilibrium bulk curve ($\tau^{eq}$ vs. $1/T$). To obtain the other (colored) solid lines, we select 5 points from the dashed lines (nonequilibrium extrapolations) for each film thickness. These points are chosen to span +/− one decade in relaxation time around the typical experimental 100-second timescale for $T_g$, with a resolution of ½ decade, which we choose to match the typical timescale range and resolution of cooling-rate $T_g$ experiments that have yielded similar data. Then, we fit these five points to an Arrhenius functional form, which provides a reasonable fit over this limited temperature range. The colored solid lines represent the extrapolation of these Arrhenius fits to higher and lower temperatures.

**Supplementary Figures and Tables**

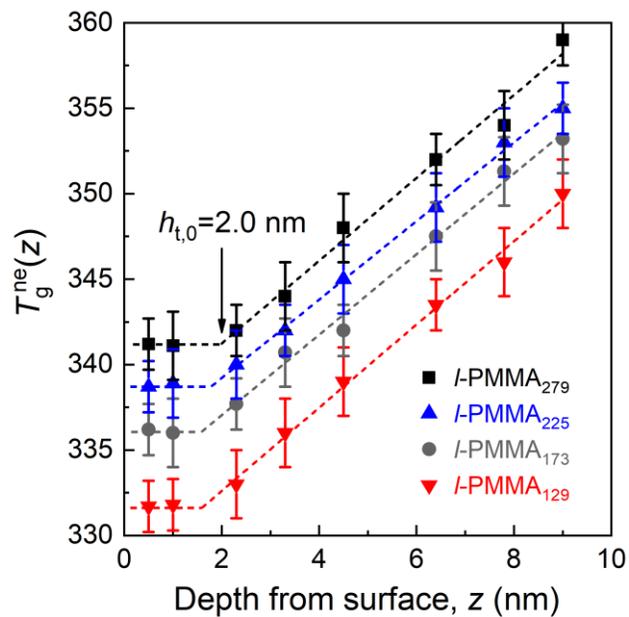

**Fig. S1. Experimental depth profiles $T_g^{ne}(z)$ taken at the surface of *l*-PMMA polymers with degrees of polymerization of 279, 225, 173, and 129.** Dashed lines are the best fits to a linear gradient.

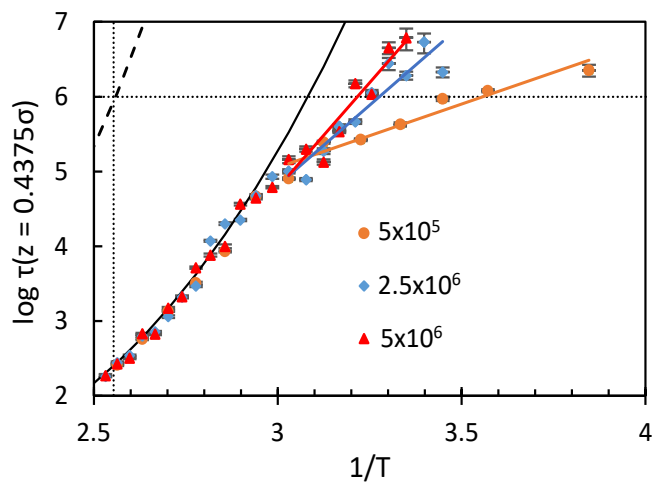

**Fig. S2. Simulated surface relaxation times (z = 0.4375 σ) with longer initial chain conformational annealing periods but same maximum segmental annealing time—equivalent of Fig. 2.** The orange points correspond to those shown in the main text, while the other data sets correspond to longer annealing periods. The solid black curves represent fits to the equilibrium data, as described in Supplementary Simulation Methods. The straight lines in colors matching the data sets are Arrhenius fits to the data deviating from the equilibrium curve. Systematic downward deviations of these data from the equilibrium curve at large τ indicate nonequilibrium effects, which are not removed by longer chain conformational annealing. The dashed line represents the equilibrium line for bulk polymer.

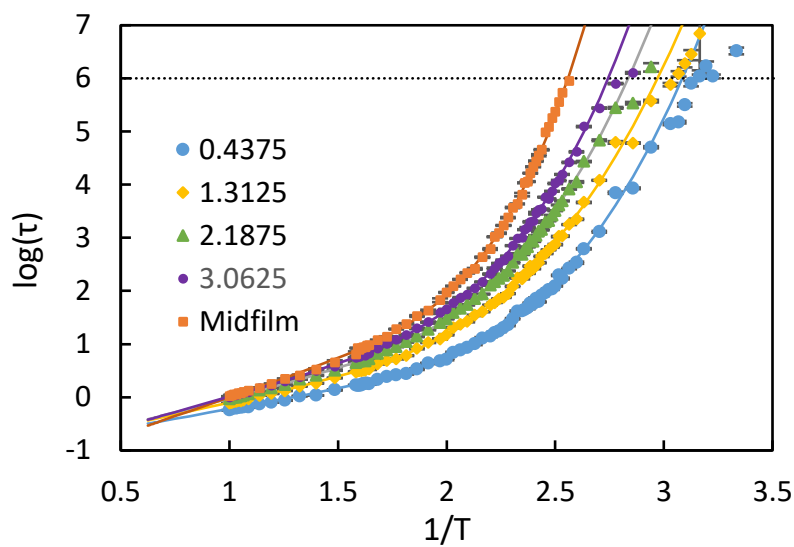

**Fig. S3. Figure equivalent to Fig. 2, after subjecting the system to a highly efficient chain-rebridging annealing protocol bringing chain configurations closer to equilibrium.** Logarithmic relaxation time (in units of Lennard-Jones time $\tau_{LJ}$) plotted against reciprocal temperature (symbols) for a 1000-bead chain film after an equivalent high-temperature annealing time of $5 \times 10^6\,\tau_{LJ}$ or longer, at various distances from the free surface (indicated in the legend). The solid lines represent fits to the equilibrium data with segmental annealing times $\leq 10^5\,\tau_{LJ}$, as described in Supplementary Simulation Method. Systematic downward deviations of the data from the equilibrium solid lines near the surface at large $\tau$ indicate nonequilibrium effects, which are not eliminated by improving the chain configurations towards equilibrium with the rebridging protocol.

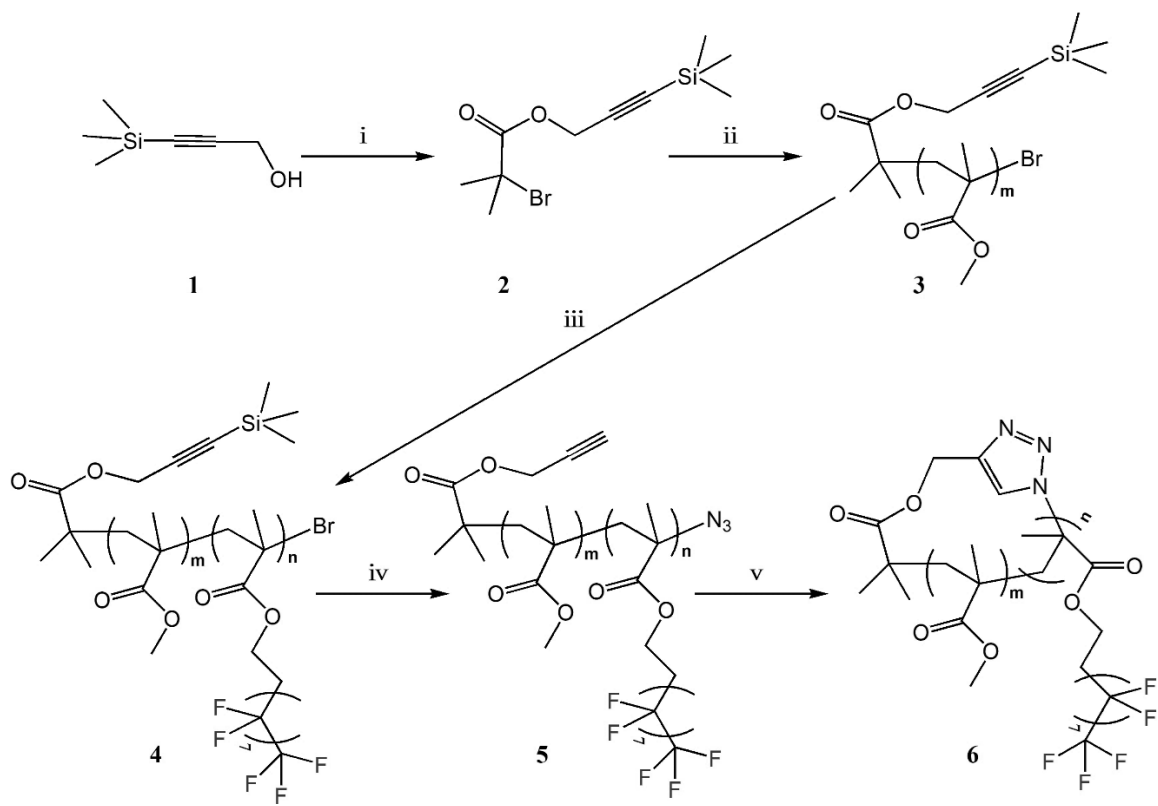

**Fig. S4. Schematic illustrations of the synthetic route for fluoro-labeled *l*-and *c*-PMMA.** The reagents and reaction conditions used in various steps of the process are: (i) THF, TEA, BIBB, 273 K; (ii) CuBr$_2$, PMDETA, MMA, 353 K; (iii) CuBr$_2$, PMDETA, MMA (a little), FMA, BTF, 385 K; (iv) a. NaN$_3$, DMF, 298 K; b. TBAF, THF, 298 K; (v) CuBr, PMDETA, toluene, high dilution, 298 K. The intermediate products are labeled **1-6**, with product 5 being *l*-PMMA$_m$-b-FMA$_n$ and product 6 being *c*-PMMA$_m$-b-FMA$_n$.

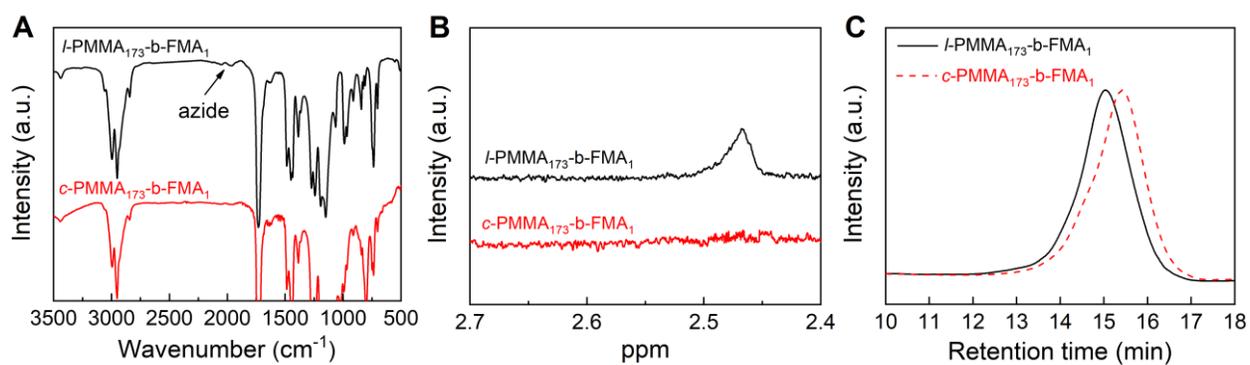

**Fig. S5. Representative characterization data of the *l*- and *c*-PMMA polymers.** (**A**) FTIR spectra (**B**) $^1$H NMR spectra and (**C**) GPC spectra. The polymers are *l*- and *c*-PMMA$_{173}$-b-FMA$_1$.

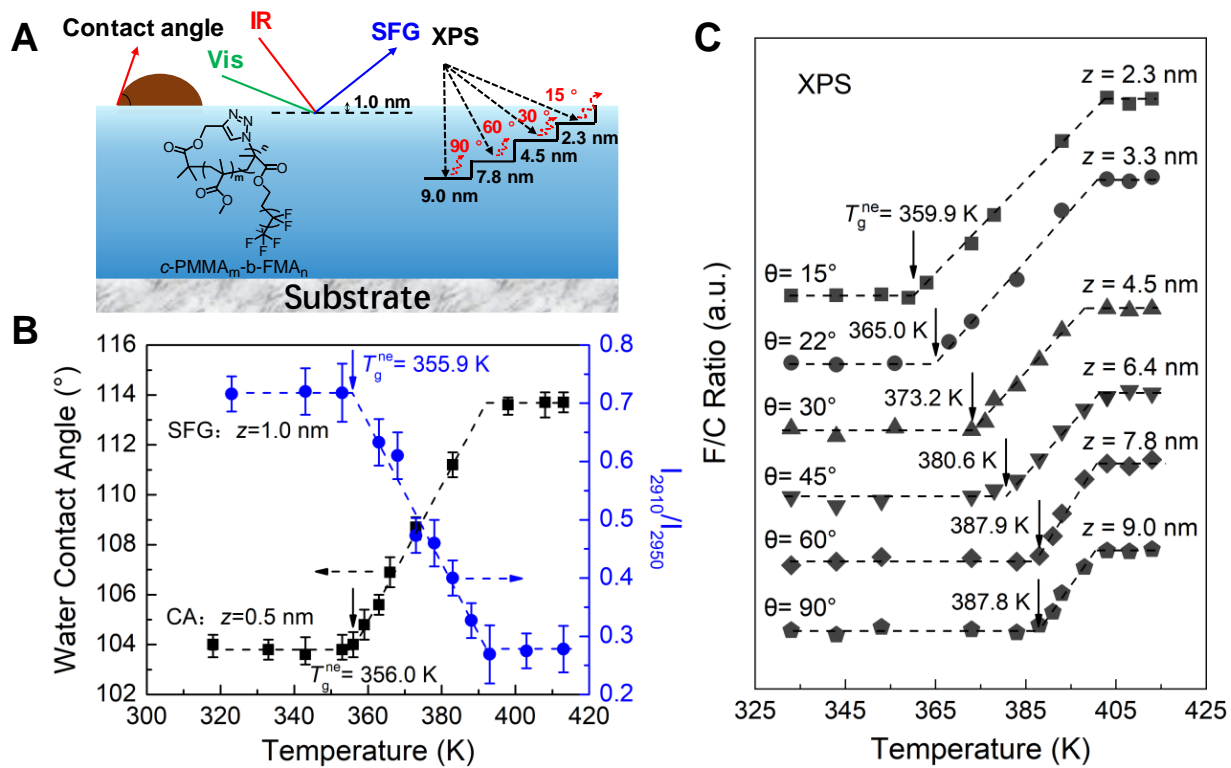

**Fig. S6. Schematic and data illustrating our method of measuring the depth profile $T_g^{ne}(z)$.** (**a**) This schematic illustrates the investigation depths of the WCA, SFG and XPS techniques used in this study, along with the chemical structures of $c$-PMMA$_m$-b-FMA$_n$. (**B**) & (**C**) Temperature dependences of the WCA and $I_{2910}/I_{2950}$ (by SFG), and the F/C ratio (by XPS) of $c$-PMMA$_{173}$-b-FMA$_1$ bulk films. The data were obtained at various take-off angles, which correspond to different probe depths, $z$.

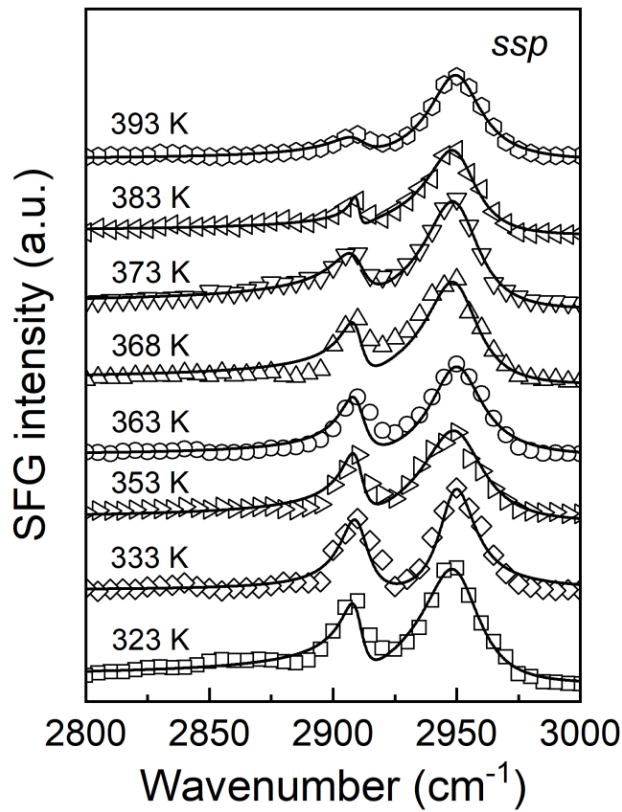

**Fig. S7. Representative SFG spectra obtained at various temperatures across $T_g$.** *ssp* SFG spectra of *c*-PMMA$_{173}$-b-FMA$_1$ obtained at various temperatures, as indicated. The probe depth was set at 1.0 nm, where the corresponding $T_g$ is 355.9 K, as shown in Fig. *1*.

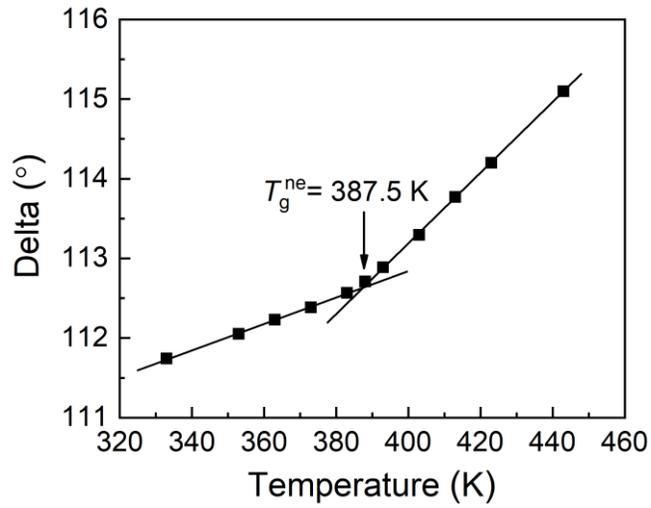

**Fig. S8. Determination of mean-film $T_g^{ne}$ using ellipsometric data.** Evolution of the ellipsometric angle, Delta, with temperature for a 106-nm-thick *c*-PMMA$_{173}$ film supported on silicon, using a heating scheme that includes a 24-h waiting time after each temperature change. Error bars are smaller than the size of the symbol.

**Table S1. Raw results of simulated density data as a function of reduced Lennard-Jones temperature**.

| Temp | Distance z from the surface | | | | | | | | | | | |
|---|---|---|---|---|---|---|---|---|---|---|---|---|
| | 0.4375 | 1.3125 | 2.1875 | 3.0625 | 3.9375 | 4.8125 | 5.6875 | 6.5625 | 7.4375 | 8.3125 | 9.1875 | 10.0625 |
| **1.600** | 0.680 | 0.900 | 0.961 | 0.971 | 0.973 | 0.973 | 0.974 | 0.974 | 0.974 | 0.974 | 0.974 | 0.972 |
| **1.580** | 0.722 | 0.917 | 0.968 | 0.975 | 0.979 | 0.978 | 0.977 | 0.977 | 0.979 | 0.978 | 0.977 | 0.976 |
| **1.560** | 0.699 | 0.912 | 0.969 | 0.981 | 0.980 | 0.984 | 0.982 | 0.981 | 0.981 | 0.981 | 0.981 | 0.982 |
| **1.540** | 0.680 | 0.912 | 0.972 | 0.985 | 0.986 | 0.988 | 0.987 | 0.986 | 0.986 | 0.987 | 0.985 | 0.984 |
| **1.520** | 0.733 | 0.934 | 0.984 | 0.990 | 0.992 | 0.991 | 0.989 | 0.990 | 0.991 | 0.990 | 0.991 | 0.989 |
| **1.500** | 0.705 | 0.925 | 0.984 | 0.993 | 0.995 | 0.996 | 0.997 | 0.996 | 0.994 | 0.996 | 0.996 | 0.993 |
| **1.480** | 0.757 | 0.948 | 0.992 | 0.999 | 0.999 | 1.000 | 0.999 | 0.999 | 1.002 | 1.001 | 0.999 | 0.999 |
| **1.460** | 0.746 | 0.954 | 0.996 | 1.003 | 1.004 | 1.004 | 1.004 | 1.003 | 1.004 | 1.001 | 1.004 | 1.002 |
| **1.341** | 0.783 | 0.991 | 1.026 | 1.029 | 1.030 | 1.027 | 1.033 | 1.029 | 1.031 | 1.027 | 1.029 | 1.026 |
| **1.223** | 0.799 | 1.016 | 1.056 | 1.058 | 1.054 | 1.055 | 1.058 | 1.052 | 1.058 | 1.056 | 1.056 | 1.055 |
| **1.104** | 0.835 | 1.053 | 1.081 | 1.081 | 1.083 | 1.081 | 1.082 | 1.081 | 1.082 | 1.079 | 1.078 | 1.089 |
| **0.986** | 0.894 | 1.088 | 1.106 | 1.109 | 1.108 | 1.110 | 1.112 | 1.113 | 1.107 | 1.109 | 1.108 | 1.109 |
| **0.867** | 0.961 | 1.120 | 1.136 | 1.137 | 1.140 | 1.137 | 1.135 | 1.135 | 1.139 | 1.136 | 1.142 | 1.132 |
| **0.749** | 1.018 | 1.151 | 1.166 | 1.163 | 1.169 | 1.166 | 1.163 | 1.162 | 1.166 | 1.164 | 1.165 | 1.163 |
| **0.630** | 1.064 | 1.178 | 1.195 | 1.200 | 1.190 | 1.195 | 1.188 | 1.198 | 1.196 | 1.195 | 1.192 | 1.201 |
| **0.611** | 1.087 | 1.186 | 1.198 | 1.201 | 1.193 | 1.202 | 1.199 | 1.201 | 1.197 | 1.196 | 1.201 | 1.196 |
| **0.591** | 1.067 | 1.185 | 1.205 | 1.205 | 1.201 | 1.202 | 1.208 | 1.204 | 1.209 | 1.202 | 1.202 | 1.205 |
| **0.571** | 1.041 | 1.196 | 1.210 | 1.208 | 1.206 | 1.214 | 1.206 | 1.208 | 1.210 | 1.209 | 1.207 | 1.209 |
| **0.552** | 1.116 | 1.203 | 1.211 | 1.214 | 1.217 | 1.216 | 1.211 | 1.214 | 1.213 | 1.213 | 1.216 | 1.216 |
| **0.532** | 1.109 | 1.206 | 1.217 | 1.216 | 1.221 | 1.220 | 1.219 | 1.215 | 1.221 | 1.218 | 1.218 | 1.220 |
| **0.512** | 1.111 | 1.207 | 1.222 | 1.230 | 1.220 | 1.224 | 1.221 | 1.223 | 1.226 | 1.228 | 1.227 | 1.222 |
| **0.503** | 1.118 | 1.214 | 1.225 | 1.226 | 1.227 | 1.228 | 1.223 | 1.229 | 1.226 | 1.229 | 1.224 | 1.228 |
| **0.494** | 1.159 | 1.221 | 1.229 | 1.228 | 1.233 | 1.221 | 1.232 | 1.225 | 1.229 | 1.227 | 1.231 | 1.230 |
| **0.485** | 1.161 | 1.222 | 1.229 | 1.234 | 1.233 | 1.230 | 1.230 | 1.233 | 1.229 | 1.232 | 1.235 | 1.229 |
| **0.476** | 1.160 | 1.222 | 1.233 | 1.230 | 1.235 | 1.238 | 1.234 | 1.233 | 1.235 | 1.233 | 1.233 | 1.234 |
| **0.467** | 1.151 | 1.226 | 1.237 | 1.238 | 1.239 | 1.234 | 1.234 | 1.238 | 1.235 | 1.235 | 1.237 | 1.235 |
| **0.458** | 1.149 | 1.228 | 1.238 | 1.239 | 1.236 | 1.237 | 1.243 | 1.237 | 1.240 | 1.236 | 1.237 | 1.240 |
| **0.449** | 1.144 | 1.231 | 1.241 | 1.239 | 1.241 | 1.239 | 1.239 | 1.244 | 1.237 | 1.238 | 1.240 | 1.242 |
| **0.446** | 1.134 | 1.229 | 1.240 | 1.242 | 1.240 | 1.243 | 1.243 | 1.241 | 1.243 | 1.242 | 1.239 | 1.241 |
| **0.442** | 1.159 | 1.234 | 1.241 | 1.246 | 1.241 | 1.240 | 1.244 | 1.244 | 1.240 | 1.247 | 1.241 | 1.245 |
| **0.438** | 1.155 | 1.230 | 1.245 | 1.245 | 1.246 | 1.243 | 1.247 | 1.240 | 1.246 | 1.244 | 1.243 | 1.242 |
| **0.435** | 1.205 | 1.236 | 1.246 | 1.248 | 1.247 | 1.247 | 1.243 | 1.243 | 1.244 | 1.244 | 1.248 | 1.245 |
| **0.431** | 1.210 | 1.238 | 1.246 | 1.245 | 1.245 | 1.250 | 1.251 | 1.244 | 1.246 | 1.246 | 1.247 | 1.245 |
| **0.427** | 1.205 | 1.239 | 1.249 | 1.248 | 1.248 | 1.245 | 1.250 | 1.247 | 1.245 | 1.244 | 1.256 | 1.243 |
| **0.425** | 1.208 | 1.237 | 1.251 | 1.248 | 1.248 | 1.251 | 1.248 | 1.246 | 1.247 | 1.246 | 1.252 | 1.244 |
| **0.422** | 1.202 | 1.238 | 1.251 | 1.246 | 1.250 | 1.249 | 1.246 | 1.247 | 1.249 | 1.250 | 1.251 | 1.244 |
| **0.419** | 1.201 | 1.241 | 1.248 | 1.252 | 1.252 | 1.248 | 1.249 | 1.245 | 1.249 | 1.252 | 1.250 | 1.249 |
| **0.417** | 1.204 | 1.242 | 1.249 | 1.253 | 1.249 | 1.250 | 1.249 | 1.253 | 1.247 | 1.250 | 1.249 | 1.251 |
| **0.414** | 1.209 | 1.240 | 1.251 | 1.253 | 1.253 | 1.249 | 1.252 | 1.249 | 1.249 | 1.248 | 1.251 | 1.249 |
| **0.412** | 1.193 | 1.242 | 1.251 | 1.256 | 1.249 | 1.253 | 1.250 | 1.249 | 1.253 | 1.250 | 1.254 | 1.249 |
| **0.409** | 1.197 | 1.243 | 1.254 | 1.246 | 1.254 | 1.252 | 1.251 | 1.254 | 1.253 | 1.253 | 1.255 | 1.248 |
| **0.407** | 1.196 | 1.242 | 1.252 | 1.254 | 1.253 | 1.253 | 1.253 | 1.254 | 1.250 | 1.256 | 1.252 | 1.250 |
| **0.390** | 1.192 | 1.244 | 1.255 | 1.259 | 1.256 | 1.261 | 1.261 | 1.250 | 1.262 | 1.257 | 1.255 | 1.256 |
| **0.380** | 1.205 | 1.249 | 1.260 | 1.266 | 1.264 | 1.263 | 1.253 | 1.256 | 1.259 | 1.263 | 1.258 | 1.251 |
| **0.370** | 1.263 | 1.259 | 1.265 | 1.275 | 1.265 | 1.267 | 1.263 | 1.256 | 1.243 | 1.273 | 1.250 | 1.259 |
| **0.360** | 1.186 | 1.256 | 1.266 | 1.259 | 1.271 | 1.265 | 1.269 | 1.250 | 1.266 | 1.257 | 1.270 | 1.248 |
| **0.350** | 1.255 | 1.263 | 1.264 | 1.275 | 1.269 | 1.275 | 1.255 | 1.258 | 1.262 | 1.267 | 1.260 | 1.253 |
| **0.340** | 1.256 | 1.263 | 1.265 | 1.274 | 1.278 | 1.274 | 1.249 | 1.265 | 1.257 | 1.271 | 1.257 | 1.265 |
| **0.330** | 1.260 | 1.264 | 1.271 | 1.283 | 1.275 | 1.263 | 1.261 | 1.266 | 1.257 | 1.269 | 1.262 | 1.261 |
| **0.320** | 1.249 | 1.276 | 1.272 | 1.276 | 1.278 | 1.262 | 1.263 | 1.268 | 1.262 | 1.278 | 1.253 | 1.274 |
| **0.310** | 1.243 | 1.266 | 1.274 | 1.271 | 1.270 | 1.263 | 1.274 | 1.270 | 1.256 | 1.261 | 1.269 | 1.265 |
| **0.300** | 1.251 | 1.278 | 1.279 | 1.261 | 1.274 | 1.265 | 1.282 | 1.258 | 1.258 | 1.262 | 1.279 | 1.269 |
| **0.290** | 1.229 | 1.282 | 1.275 | 1.269 | 1.263 | 1.275 | 1.268 | 1.276 | 1.258 | 1.276 | 1.270 | 1.272 |
| **0.280** | 1.247 | 1.278 | 1.282 | 1.265 | 1.271 | 1.273 | 1.271 | 1.267 | 1.258 | 1.269 | 1.284 | 1.263 |
| **0.270** | 1.248 | 1.278 | 1.282 | 1.262 | 1.281 | 1.270 | 1.272 | 1.274 | 1.258 | 1.274 | 1.278 | 1.271 |
| **0.260** | 1.231 | 1.275 | 1.281 | 1.269 | 1.275 | 1.275 | 1.273 | 1.273 | 1.260 | 1.268 | 1.280 | 1.271 |
| **0.240** | 1.250 | 1.275 | 1.284 | 1.264 | 1.278 | 1.275 | 1.270 | 1.271 | 1.269 | 1.262 | 1.284 | 1.264 |
| **0.230** | 1.233 | 1.269 | 1.291 | 1.262 | 1.278 | 1.268 | 1.278 | 1.279 | 1.258 | 1.272 | 1.288 | 1.269 |

**Table S2. Summary of fit results for $T_g^{ne}(z)$ data in Fig. *1* using equation S1.**

| Sample | $T_{g,bulk}^{ne}$ (K) | $T_g^{ne}(0)$ (K) | $h_{t,0}$ (nm) | $h_t$ (nm) | $[T_{g,bulk}^{ne} - T_g^{ne}(0)] / [h_t - h_{t,0}]$ (K/nm) |
|---|---|---|---|---|---|
| *c*-PMMA$_{129}$ | 387.1 | 352.1 ± 2.0 | 1.0 ± 0.2 | 7.6 ± 0.7 | 5.32 ± 0.31 |
| *c*-PMMA$_{173}$ | 387.5 | 356.0 ± 2.0 | 1.5 ± 0.2 | 7.7 ± 0.8 | 5.13 ± 0.43 |
| *c*-PMMA$_{225}$ | 388.0 | 362.8 ± 1.5 | 1.6 ± 0.2 | 7.9 ± 0.8 | 4.00 ± 0.21 |
| *c*-PMMA$_{279}$ | 388.0 | 369.9 ± 1.5 | 1.8 ± 0.2 | 7.8 ± 0.8 | 2.99 ± 0.10 |

**Table S3. Material characteristics of the polymers used in this experiment**.

| Sample | $M_n$ (kg/mol) | PDI | $T_{g,bulk}^{ne}$ | $R_g$ (nm) |
|---|---|---|---|---|
| $c$-PMMA$_{129}$-b-FMA$_1$ | 12.9 | 1.20 | 393.8 | 2.26 |
| $c$-PMMA$_{173}$-b-FMA$_1$ | 17.3 | 1.18 | 393.4 | 2.62 |
| $c$-PMMA$_{225}$-b-FMA$_2$ | 22.5 | 1.16 | 393.8 | 2.99 |
| $c$-PMMA$_{279}$-b-FMA$_2$ | 27.9 | 1.18 | 394.2 | 3.33 |
| $l$-PMMA$_{129}$-b-FMA$_1$ | 12.9 | 1.18 | 388.9 | 3.20 |
| $l$-PMMA$_{173}$-b-FMA$_1$ | 17.3 | 1.16 | 392.1 | 3.71 |
| $l$-PMMA$_{225}$-b-FMA$_1$ | 22.5 | 1.17 | 393.9 | 4.23 |
| $l$-PMMA$_{279}$-b-FMA$_1$ | 27.9 | 1.15 | 393.9 | 4.71 |